# Black holes at work

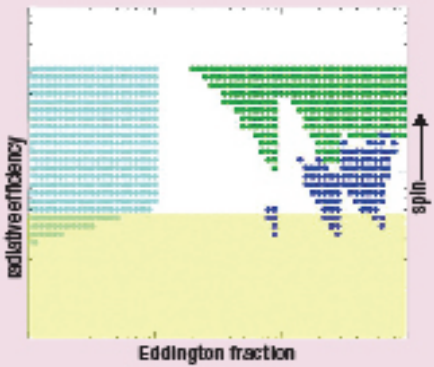

1: The radiative efficiency of AGN (green, unabsorbed; blue absorbed) populations.

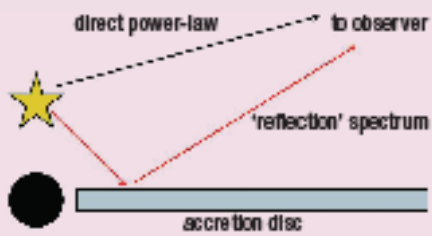

2: The power law continuum in AGN comes to us directly and reflected off the disc.

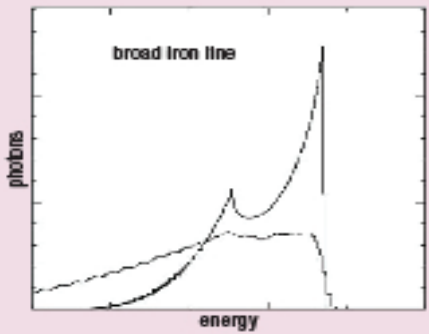

3: Strong gravity distorts the iron K line. The broader line is from a fast spinning black hole.

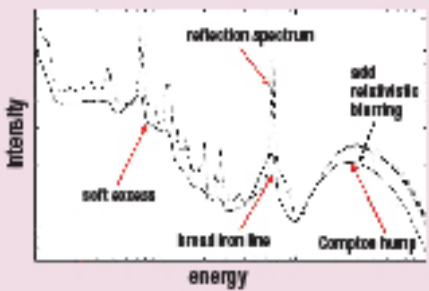

4: Reflection spectrum expected (dashed); observed (solid after relativistic blurring).

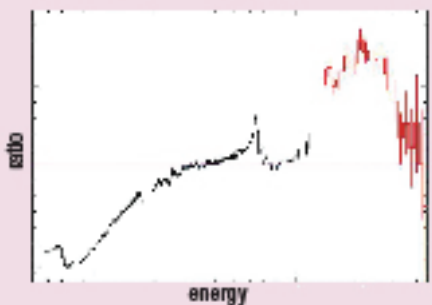

5: The 0.5–50 keV Suzaku spectrum of MCG–6–30–15. (Miniutti 2007)

Andrew C Fabian reviews the role that black holes play in the evolution of the galaxies that surround them, in his 2009 Presidential Address to the Royal Astronomical Society.

**ABSTRACT**

Massive black holes are ubiquitous, occurring at the centres of all massive galaxies and possibly many low mass ones. They are no ornament which just happens to be there, but play a role vital to the growth and structure of the host galaxy. How they do this has come to be known as cosmic feedback and how it works, indeed how black holes work, is the subject of this paper.

The visible sky is dominated by objects powered by nuclear fusion such as stars and galaxies. Shifting to shorter wavelengths in the X-ray band reveals an extragalactic sky powered by gravity: gravitational energy released by matter falling into black holes. All matter is in motion and as it falls towards an object as small as a black hole most matter has sufficient angular momentum to go into orbit about the hole, so forming an accretion disc. Here, viscosity due to magnetic turbulence causes the angular momentum to be transferred outward as the mass accretes inward. When the accretion rates are high, considerable amounts of gravitational energy are released as radiation, and in some circumstances as powerful jets.

The radiative efficiency of accretion, $\eta$, defined through the radiative luminosity $L = \eta \dot{M} c^2$, for a black hole smoothly varies from 5.7% for no spin (Schwarzschild case) to 32% for the maximum plausible spin (Thorne 1974). This compares with 0.7% for complete nuclear fusion of hydrogen and $10^{-10}$ for chemical reactions. The energy content of hydrogen is almost a billion times greater when used in accretion than when used as a rocket fuel. At low mass flow rates, accretion may become radiatively inefficient in a manner and regime which does not immediately concern us here.

We can get some idea of the mean accretion efficiency of luminous black holes in the universe by applying an argument published by Soltan (1982). This uses the familiar mass and energy equation $E = mc^2$ in its density form using the energy density in radiation, due to accretion, $\varepsilon(1+z) = \eta/(1-\eta)\rho_{BH} c^2$, where $\rho_{BH}$ is the mean mass density in massive black holes. It is independent of the details of the cosmology except for the use of the mean redshift $z$, since photons are redshifted but matter is not. Energy density $\varepsilon$ can be measured from, say, the X-ray background, which is known to be due to accreting black holes, or quasars and active galaxies selected in some other way. Bolometric corrections and adjustments for absorption etc are needed to go from whatever observations are used to the total radiation released. The result from many different approaches indicate that $\eta \sim 0.1$ (e.g. Yu and Tremaine 2002, Elvis *et al.* 2002, Marconi *et al.* 2004).

We have recently carried out an analysis (fig. 1, Raimundo and Fabian 2009) in which the active galactic nuclei (AGN) population was divided into three separate populations according to the level of absorption in the host galaxy and to the Eddington fraction ($\lambda$: the ratio of the bolometric luminosity to the Eddington limit for that black hole mass). High $\varepsilon$ is required for most of the sources, especially the unobscured ones. (In practice there are many more than three subclasses of AGN. A large fraction of the black

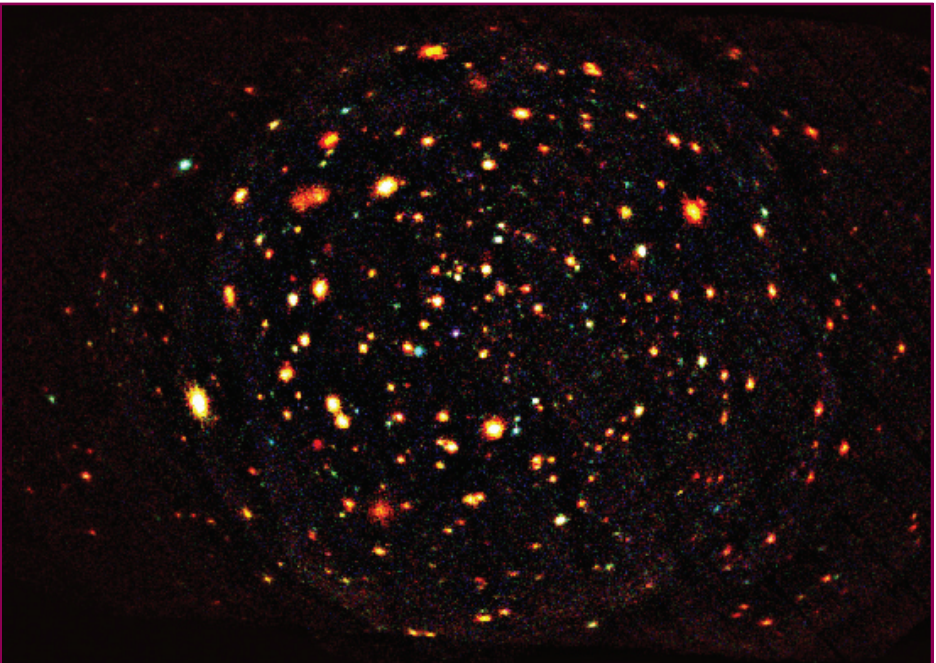

6: XMM-Newton image of the Lockman Hole, 30 arcmin across. Red, green and blue denote softer to harder X-rays. (ESA/NASA)

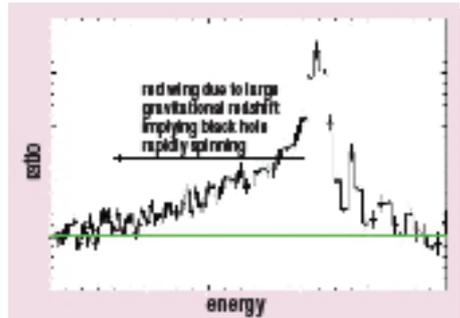

7: The broad iron line in the 3–8 keV spectrum of MCG–6-30-15. (Miniutti *et al.* 2007)

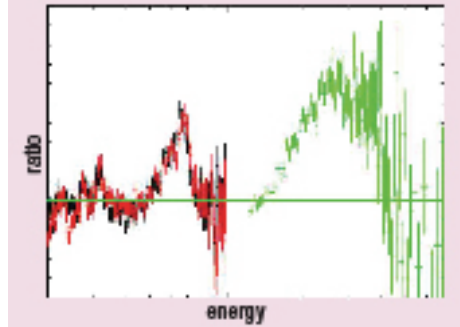

8: The 2–70 keV Suzaku spectrum of GX339-4. (J Miller *et al.* 2008)

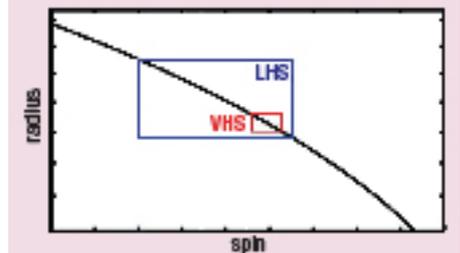

9: Constraints on inner disc radius and spin for GX339-4 in the Low Hard and Very High States. (Reis *et al.* 2008)

holes require $\varepsilon$ to be high (>0.1) which means that the dimensionless spin $a = cJ/GM^2 > 0.67$, where $J$ is the angular momentum of the black hole.)

Massive black holes acquire spin by accretion from the surrounding accretion disc and ultimately from the angular momentum of the accreting gas. If they double their mass accreting from gas with a roughly constant angular momentum direction then the end result is high spin. In contrast, it is low spin if the direction is more sporadic and random. The distribution of spins is therefore a source of information of the accretion history (Volonteri *et al.* 2005, King and Pringle 2007).

### Black holes in close-up

While the mass of a black hole can be measured by observation of matter orbiting at some considerable distance (even thousands of gravitational radii) from the hole, the measurement of spin requires much closer inspection. The test particles observed essentially need to be within 6 gravitational radii ($6r_g = 6GM/c^2$). The power output rises fivefold as the spin goes from 0 to 0.998, so more than 75% of the luminosity from accretion onto a rapidly spinning hole originates from with $6r_g$ and more than 50% originates from within $2.5r_g$.

In order to understand how accreting black holes work we need some diagnostics of the innermost regions, where most of the power emerges, and also where the gas is hottest. The X-ray spectrum can give some powerful diagnostics though the reflection spectrum (fig. 2). This is the backscattered emission produced when the power law continuum, which is produced by Comptonization of thermal emission from the disc by hot coronal electrons, illuminates the rest of the accretion flow. It consists of a flat continuum superposed with emission lines, especially those due to iron-K, and a broad hump around 20–30 keV known as the Compton hump (Lightman and White 1989, George and Fabian 1991). As the intensity of the irradiation rises, then surface layers become progressively ionized, leading to changes in the reflection spectrum (Ross and Fabian 1993, 2005). When the accretion flow is a disc then the reflection spectrum appears blurred in a characteristic manner (figs 3–7, Fabian *et al.* 1989,

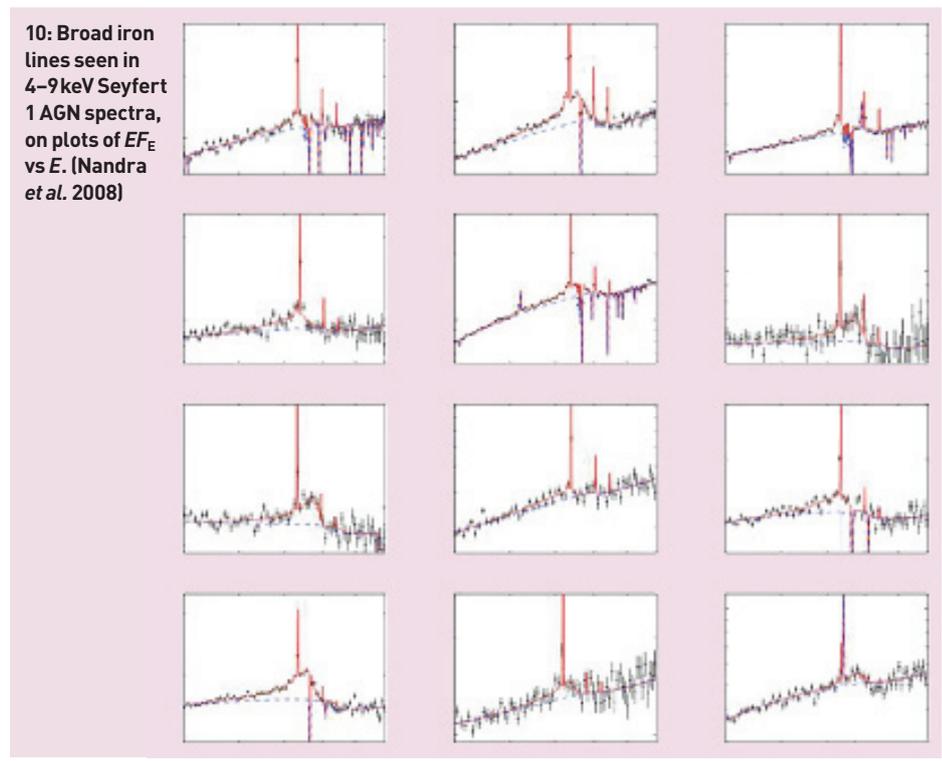

10: Broad iron lines seen in 4–9 keV Seyfert 1 AGN spectra, on plots of $EF_E$ vs $E$. (Nandra *et al.* 2008)

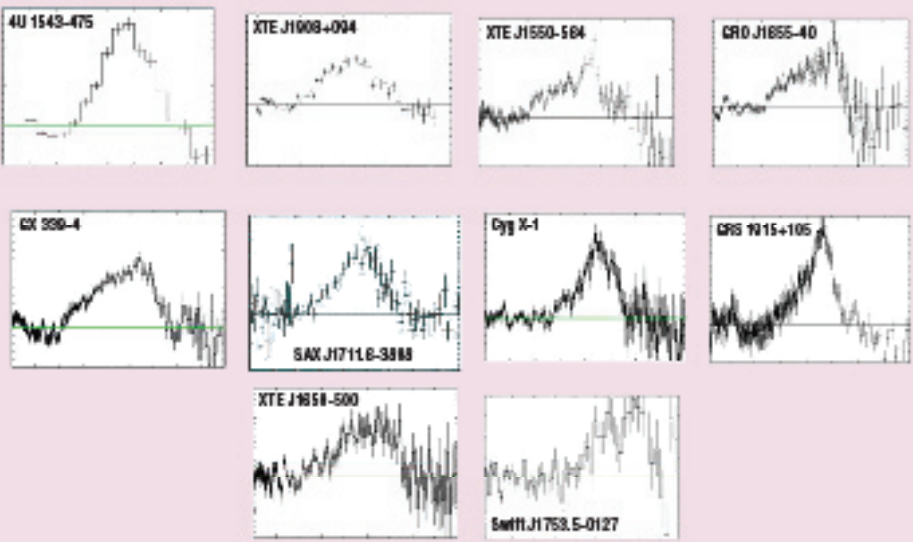

11: Broad iron lines seen in 2–10 keV GBH spectra. (Courtesy J M Miller)

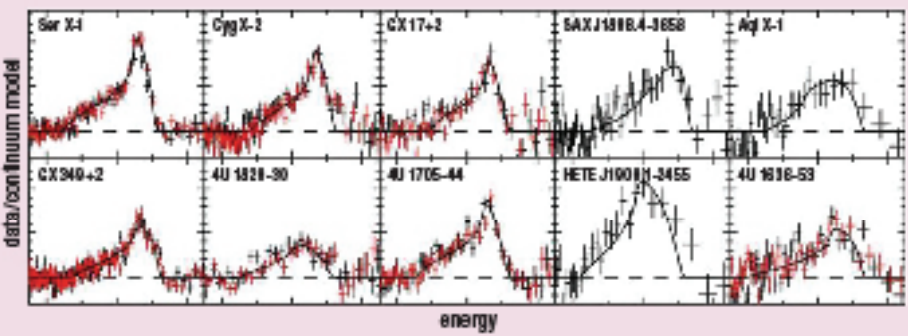

12: Broad iron lines seen in accreting neutron stars. (Cackett *et al.* 2008)

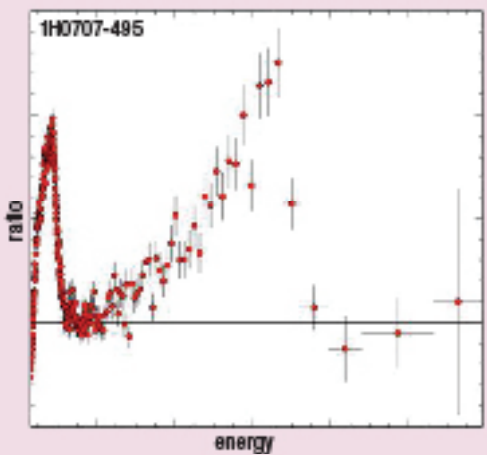

13: X-ray light curve, below, and 0.4–12 keV spectrum, left, (shown as a ratio to a power law continuum) of the Seyfert galaxy 1H0707-495. (Fabian *et al.* 2009b)

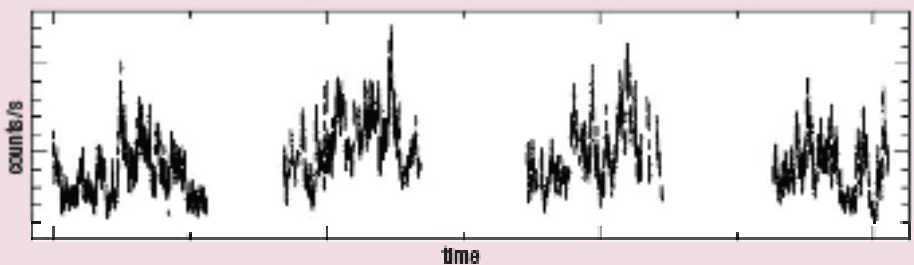

Laor 1991). The dominant feature in the well-observed 2–10 keV region is the broad iron line, the high energy wing of which depends mostly on the disc inclnation and the low energy wing depends mostly on the degree of gravitational redshift, i.e. the inner radius of the disc.

The inner radius of the accretion disc depends on the spin of the hole arising from dragging of inertial frames in the Kerr metric. It is $6r_g$ for a non-spinning hole and drops in to approach $1r_g$ at the highest spin. At this point the iron line, which has a rest energy in the range 6.4–6.9 keV, extends down to 2–3 keV. A further important relativistic effect for reflection around rapidly spinning holes is that of light bending. Emission from the corona is bent down towards the disc and black hole, making the reflection much stronger than it would be in simple Euclidean space.

Strong reflection effects are common in the X-ray spectra of AGN. Many show a soft excess which can be characterized as blackbody emission, but all show approximately the same temperature, 250 eV. If due to reflection, this is just the blurred conglomeration of soft line emission and bremsstrahlung from the irradiated disc surface. As already mentioned, a broad iron line appears in the 2–10 keV band and a Compton hump is often seen peaking at 30 keV.

A clear example of a strong, broad iron line and Compton hump is provided by MCG–6-30-15 (Tanaka *et al.* 1995, Guainazzi *et al.* 1999, Fabian *et al.* 2002, Miniutti *et al.* 2007). Indeed the broad hump is so large that it demands a strong iron emission line to be present. For a standard geometry, the only way it can agree with the observed spectrum is if that line is strong blurred. Absorption interpretations that attempt to mimic the broad iron line feature using absorption edges (e.g. L Miller *et al.* 2008) are unable to match the whole spectrum unless the absorber lies exactly along our line of sight and only subtends an implausible few percent of the sky at the source (Reynolds *et al.* 2009). The width of the broad iron line in MCG–6-30-15, and indeed the degree to which its whole reflection spectrum needs to be relativistically broadened, supports the view that the black hole there is a rapidly spinning Kerr hole with a dimensionless spin parameter $a > 0.98$ (Brenneman and Reynolds 2008, see also Dabrowski *et al.* 1997).

Broad iron lines are commonly, but not always, seen in deep X-ray spectra of bright non-jetted AGN (Guainazzi *et al.* 2007, Nandra *et al.* 2008, fig. 10). They are also common in the X-ray spectra of galactic black hole (GBH) sources (fig. 11, Miller 2007 and references therein) and accreting neutron stars (fig. 12 Cackett *et al.* 2009). The Suzaku X-ray spectrum of the GBH source GX 339-4, for example, shows a clear broad line and Compton hump (fig. 8, J Miller *et al.* 2008). Suzaku and XMM data in two different source states show that the spin of that black hole $a \sim 0.93$ (fig. 9, Reis *et al.* 2008). Other GBH sources show a broad distribution in spin (Miller *et al.* 2009). Complementary work on GBH spin which models the thermal emission in the high state (Shafee *et al.* 2006) also shows a range of spin parameters.

Relativistically broadened lines appear to be a common feature of accreting black holes. In any individual case such broad features may be mimicked by absorption features, especially if systematic uncertainties are introduced together with no limit to the number of absorbers. However, while broad emission features are black hole mass invariant, the absorption properties are not. It is much more difficult to explain away the broad lines in GBH than for AGN. Conversely, if they are seen in GBH, why would they be missing in AGN?

We have recently obtained 500 ks of XMM data on the (optically classified) narrow-line Seyfert 1 galaxy 1H0707-495 (fig. 13, Fabian *et al.* 2009b). The light curve shows an enormous level of rapid variability. The X-ray spectrum is steep and can be characterized as a power-law continuum with two superimposed broad lines (fig. 14). The lines are due to Fe-K and -L shell transitions and show the same level of relativistic blurring. The reason that we see both lines in this object is due to its apparent very high iron abundance (between 5 and 10 times the solar value). The Fe-L line also shows a hint of O emission in the red (low energy) wing. Using a full reflection model we find that the inferred spin is again high ($a > 0.98$).

The spectral variability of the source can be decomposed into a highly variable power law continuum and a less variable reflection component (fig. 15). Above 1 keV the power law continuum dominates; below 1 keV reflection dominates. This is exciting since we can use the bright soft X-ray band to search for reverberation: the time delay expected when the reflecting component appears to lag behind in response to changes in the continuum due to the light travel time. Such reverberation lags are expected to be on the timescale of 10s of seconds (the mass of the black hole in 1H0707-495 is expected to be a few million solar masses and the light crossing time of the gravitational radius of $10^6 M_\odot$ is 5 s).

The data do show a significant lag of about 30 s for variations on timescales less than about 15 minutes (figs 16, 17). On longer timescales there is a positive lag, which is likely due to ionization changes in the disc following mass flow rate changes. The short timescale lag is the first time that reverberation has been seen from the innermost disc immediately around the black hole.

Absorption models have been discussed for 1H0707-495, but it is very difficult to see how they can produce two apparent broad emission lines and such a short timescale lag. Any deep Fe-L absorption edge used to mimic the blue (high energy) wing of an Fe-L emission line also requires deep iron UTA absorption around 0.75 keV, which is not seen.

The above discussion shows that we do have a powerful diagnostic in the reflection signature from the inner accretion flow. Much power originates in a compact central corona only a few gravitational radii above the black hole. Strong light bending is mandatory there, which means that the reflection can appear to exceed the power-law component and the variability become confusing (see Miniutti and Fabian 2004 and Fabian and Miniutti 2009 for discussions). In objects with high spin, 75% of all power emerges from within $6r_g$ where strong gravity effects dominate. We are beginning to see how the process works in some detail, but much remains to be done. In particular we do not yet have good diagnostics of how jets form.

I now wish to turn from the innermost parts of the accretion flow to look at the effects of the accretion power on the bulge of the surrounding host galaxy of mass $M_{gal}$, which can be considerable. The binding energy of the galaxy is of order $M_{gal}\sigma^2$ which corresponds to $10^3 M_{BH}\sigma^2$, given that the typical central black hole has a mass $M_{BH} \approx M_{gal}/1000$. The energy released by the black hole while building its mass through accretion is $0.1 M_{BH} c^2$ or $10^{-4}(c/\sigma)^2 \approx 100$ times the binding energy of the bulge, if the velocity dispersion of the stars, $\sigma \sim 300$ kms$^{-1}$.

This is a profound conclusion. Energetically, the growth of the central black hole can destroy the galaxy. There are of course a few factors missing from the above estimate, but being conservative we can say that the black hole produces 30 times more energy than the binding energy of the host bulge (fig. 19). Of course it can only affect the bulge if there is some strong coupling between the energy flowing outward and the matter in the bulge. It is most unlikely to have any significant effect on the stars there. What the accreting black hole can do is to blow away the gas in the galaxy and thereby prevent star formation, coincidentally starving itself of fuel (Silk and Rees 1998, Fabian 1999).

There are two major modes for the interaction: one is the radiative (or quasar or wind) mode where it is the radiation from the AGN which couples with the gas, the other is the kinetic (or jet or radio) mode where jets from AGN mechanical heat or push the gas out. The radiative mode works best on cold gas and the kinetic one on hot gas.

One approach for the radiative mode which I have been pursuing is to see how the Eddington limit is change for dusty gas, rather than hydrogen, as assumed in the standard derivation of the Eddington limit. In the dusty gas case the effect depends on the spectrum. When

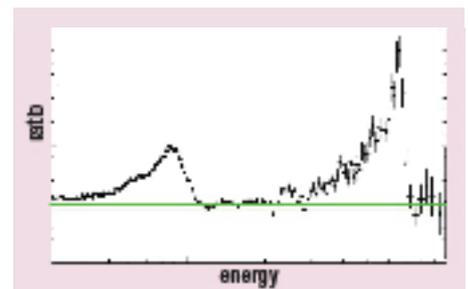

**14:** Broad iron L and K lines in 0.3–10 keV band, 1H0707-495.

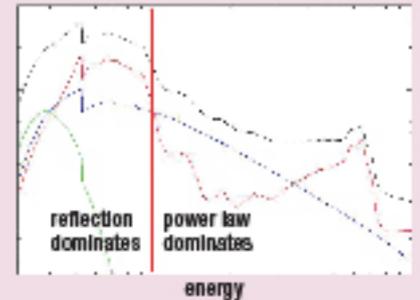

**15:** The spectra of 1H0707-495 can be modelled as highly variable power law and a less variable reflection component.

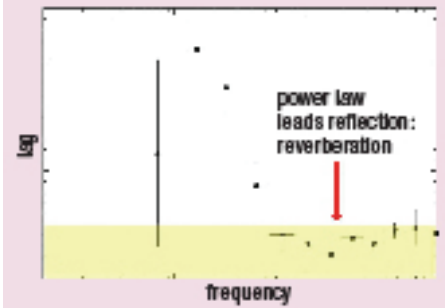

**16:** The reflection-dominated soft X-rays at <1 keV lag behind the power law dominated 1–3 keV X-rays by about 30 seconds. This is a light travel time effect (reverberation).

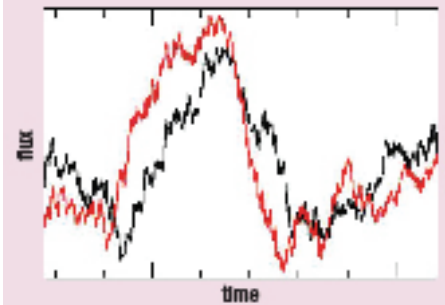

**17:** Clear example of the soft lag in this 800 s section. Red: power law; black: reflection.

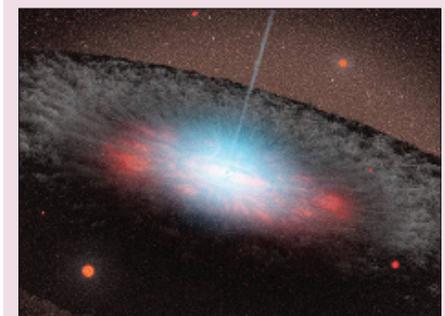

**18:** Sketch of an accreting black hole with both intense radiation and jetted emission.

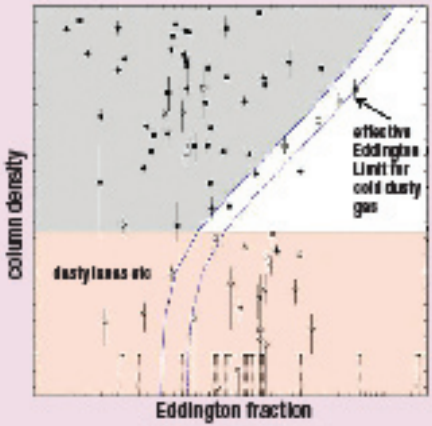

$E_{BlackHole} > 30 \times E_{Galaxy}$

$E_{BlackHole}$ is the energy released by the growth of the black hole

$E_{Galaxy}$ is the gravitational binding energy of the host galaxy

19: Energy from the accretion build-up of the central black hole greatly exceeds the gravitational binding energy of the bulge.

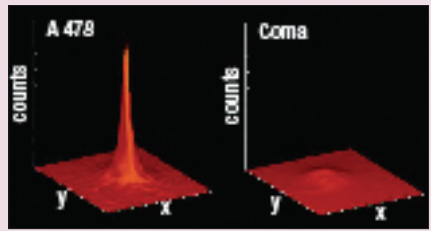

20: AGN in the Swift/BAT sample, shown as points, lie only below this effective Eddington limit. (Fabian *et al.* 2009a)

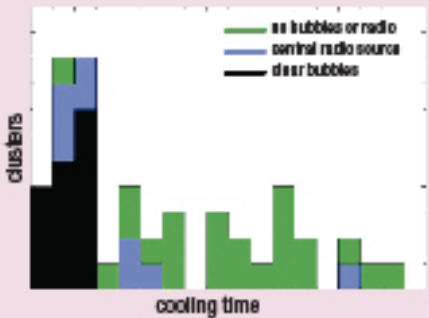

21: X-ray surface brightness profiles of a cool-core cluster (left), and a non cool core cluster (right), drawn as if observed at the same distance. (Courtesy Steve Allen)

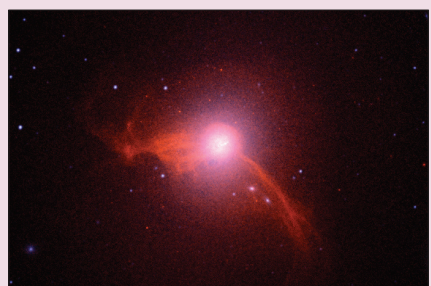

22: Histogram of bubbles and radio sources in a sample of the 55 X-ray brightest clusters. (Dunn and Fabian 2006)

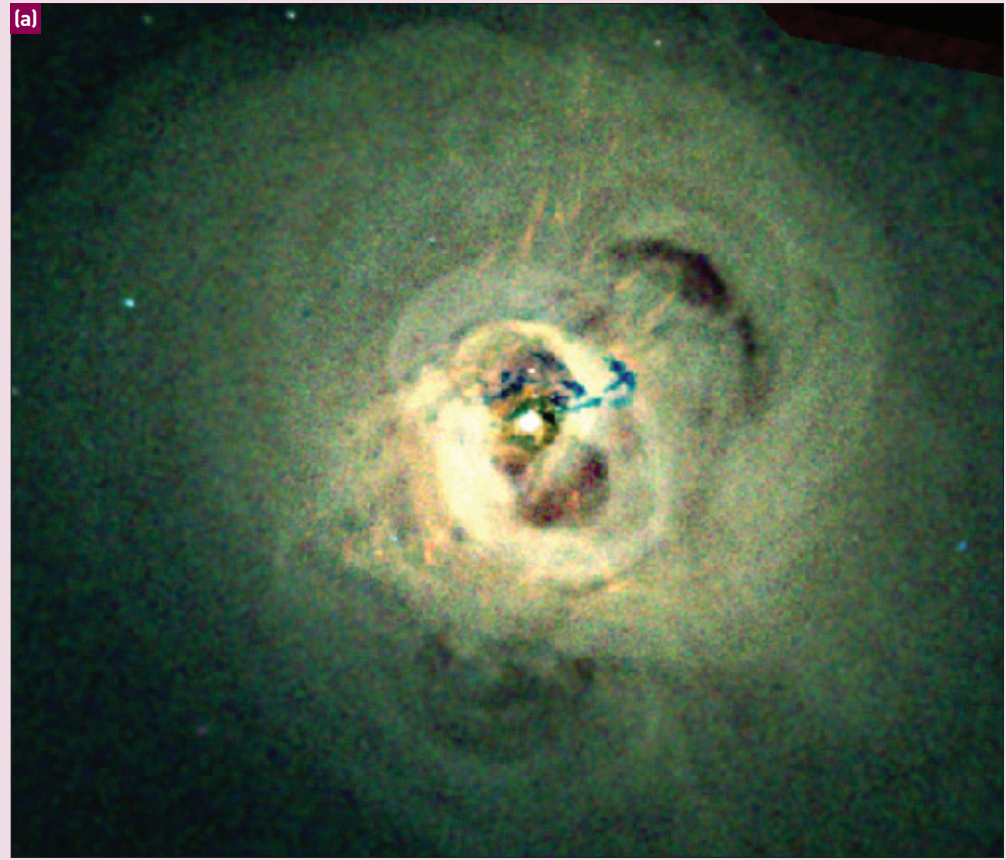

23: Chandra X-ray image of the centre of the Virgo cluster around M87, with delicate soft X-ray filaments. (Forman *et al.* 2007)

## The Perseus cluster

(a)

it is AGN-like with a large thermal peak in the far-ultraviolet, then the coupling is very strong and the effective Eddington limit reduces by over two orders of magnitude. The interaction is principally momentum driven.

The interaction on dusty gas can also be recast into a limit in the gas column density – Eddington fraction plane (fig. 20). A large triangular region at high Eddington fraction of $10^{-2}$ and above is then forbidden to long-lived gas clouds with column densities of $10^{21}$–$10^{24}\,cm^{-2}$, since the radiation there exceeds the effective Eddington limit. Interestingly, sources in the SWIFT–BAT catalogue avoid this region (Fabian *et al.* 2009a). Most of these sources are at low redshift and the result is evidence for feedback. It will be interesting to look for it around redshift 2 when feedback may have been most active.

Jets can push matter about if it happens to lie on their path. The best way, however, for a jet-driven interaction to occur is if the galaxy has a hot halo, either of its own or because it is part of a group or cluster of galaxies. Then the relativistic fluid in the jets can displace the hot gas, creating bubbles either side of the nucleus. A significant fraction of the energy in the jets can then transfer to the hot halo. If this prevents the halo from radiatively cooling to form cold clouds and stars, then the feedback terminates the stellar growth of the host galaxy.

This process is revealed in detail in X-ray images of the cores of peaked clusters of galaxies (fig. 21). About two-thirds of clusters have the X-ray emission from the diffuse intra-cluster medium peaked on the central galaxy. It means that the radiative cooling time at the centre is less than about 7 Gyr, so radiative cooling could significantly have affected the core since redshift one. In about one-third of clusters the peak is stronger, such that the cooling time is less than 3 Gyr.

In the absence of any heating balancing cooling, a cooling flow would be established leading to considerable mass cooling rates, which can be 10–100$M_\odot$ per year in a typical cluster and exceed 1000$M_\odot$ per year in the most luminous ones. The larger rates would, if turned into stars, build much larger central cluster galaxies than are seen. Chandra X-ray images of the nearer and brighter of these clusters often show bubbles near the centre where the central jetted radio source has displaced the intracluster gas. X-ray spectra, and in particular XMM-Newton reflection grating spectra, show a lack of gas less than about one-third of the bulk cluster temperature, when compared with that expected from a continuous cooling flow (Peterson and Fabian 2006). Heat is being fed into the gas to balance radiative cooling by the central accreting black hole (McNamara and Nulsen 2007).

Energetically there is no great problem since the power in the jets is about what is required. The power is estimated from the size of the bubbles and the surrounding pressure measured

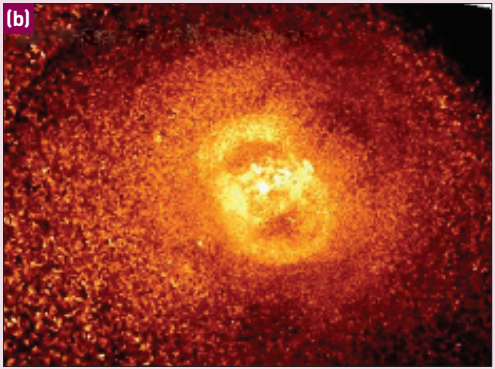
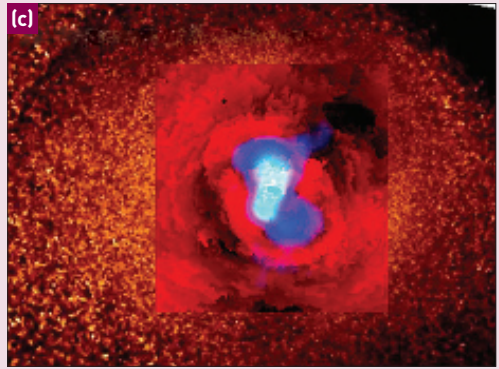
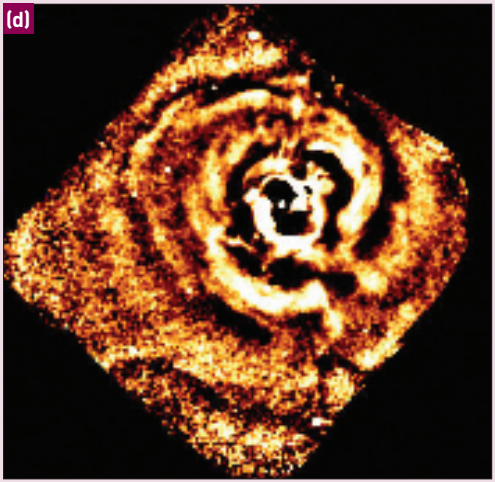

24: (a) The core of the Perseus cluster around the central galaxy NGC1275 imaged in X-rays by Chandra. (Fabian *et al.* 2006)
(b) Thermal pressure map of the region shown in (a). Note the high pressure rings around the inner bubbles created by the central radio source.
(c) As (b), but with radio emission (blue) superimposed.
(d) Unsharp masked image of the Perseus cluster core showing concentric ripples (the sound waves).

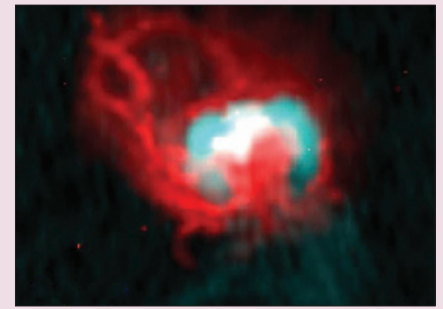

25: Unsharp, masked, Chandra X-ray image of the core of the Centaurus cluster with radio emission superimposed (blue).

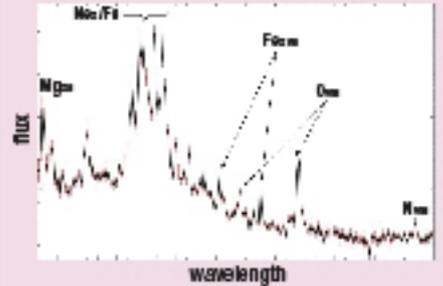

26: XMM reflection grating spectrum of cool gas (around 0.4 keV) in the core of the Centaurus cluster. (XMM/Sanders *et al.* 2007)

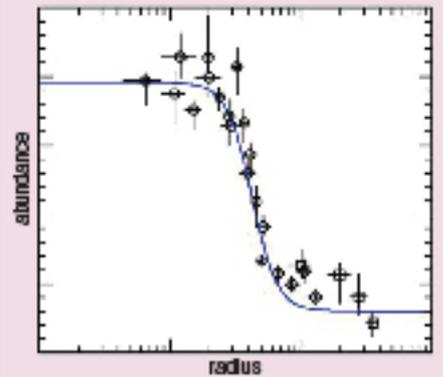

27: The steep abundance gradient of metals in the core of the Centaurus cluster.

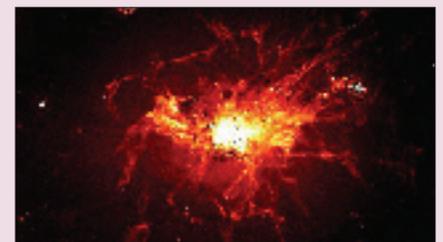

28: HST image showing the Hα and N ii line emitting filaments. (Fabian *et al.* 2008)

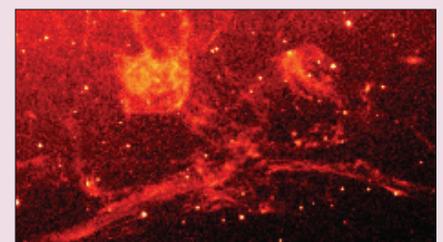

29: The spectrum of these filaments is unlike anything in our galaxy, except perhaps the Crab nebula, with low ionization and a massive molecular component.

from the X-ray data (i.e. PdV work done). The age of a bubble is obtained from buoyancy considerations. In a sample of the 55 brightest clusters (which tend to be the nearer ones and so give the best resolved measurements), 16 of the 22 with central cooling times less than 3 Gyr have clear bubbles, another 5 have bright extended central radio source and the remaining one has a weak point-like radio source (fig. 22, Dunn and Fabian 2006). This means that the duty cycle of bubbling activity is at least 70% and could be 90% or more, when projection effects are considered. Almost all regions where heating is needed do have a heat source switched on.

There are two striking points about the jet activity which were not known 10 years ago. The first is that the jets are approximately continuous (they may vary, but on a timescale of $>10^8$ yr they are continuous); the second is that the jets are highly radiatively inefficient, with the fraction of their kinetic power emerging as radiation being less than 1% and sometimes less than 0.1%. Studies of the nearest objects such as M87 show agreement with the Bondi accretion rate from the diffuse inner gas and the jet power, assuming a mass–energy efficiency of a few percent (Allen *et al.* 2006). We don't understand how most of the power from accretion is turned into a relativistic jet.

Steep abundance gradients are seen in many of the X-ray peaked clusters (fig. 27). The high abundance is due to the many supernovae in the massive central galaxy. In the nearby Centaurus cluster there is good agreement between the mass of excess iron in this region and what is expected from 8 Gyr of supernovae (Sanders *et al.* 2007). The abundance peak is a little broader than the galaxy so some diffusion is taking place, but there are no signs of any violent activity. It appears that relatively gentle bubbling must have been continuing for much of this time. Hot gas is seen (fig. 26) with a range of 10 in temperature, but the mass of the coolest components is much lower than expected if significant cooling was taking place. The central galaxy also appears red and dead. Heating and cooling appear to have been well balanced to within about 10% for the past 8 Gyr. How does such tight feedback persist?

Observations of the Perseus cluster indicate how the energy is transferred from the bubbles to the surroundings in a roughly isotropic manner (figs 24–26, Fabian *et al.* 2003, 2006). Concentric ripples in surface brightness are seen which represent ripples in pressure, or sound waves. They start as weak shocks and then spread out as sound waves, the energy flux of which agrees well with that needed to balance radiative cooling. The repeated bubbling leads to the generation of sound waves. Dissipating the energy of the sound waves depends on the viscosity and microphysics of the gas, which is of course magnetized.

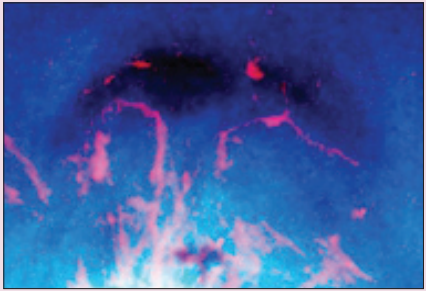
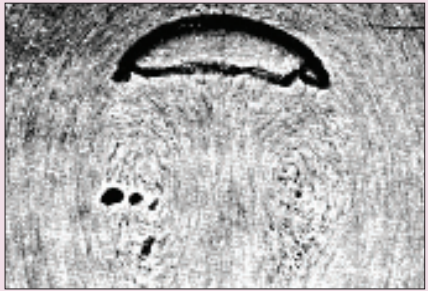

30: An outer bubble in the Perseus core (top) appears as a hole in the X-ray emission (blue) with Hα emitting filaments (pink). The filaments resemble the streamlines behind a rising cap bubble in water (bottom).

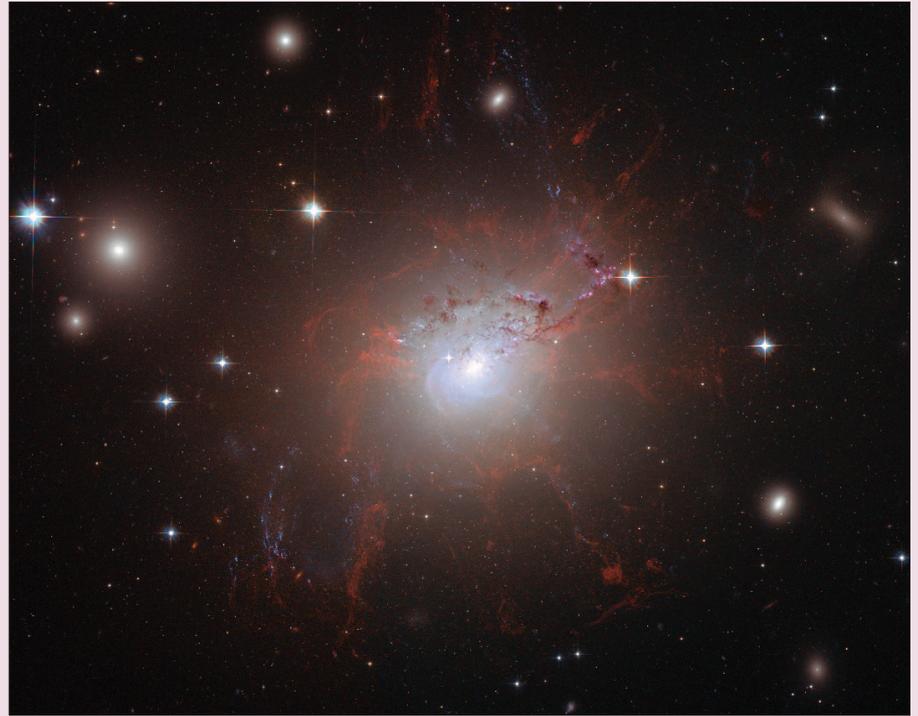

31: NGC 1275 with the Hubble Space Telescope. (Fabian *et al.* 2008)

Optical line-emitting filaments in the Perseus (fig. 28) and Centaurus clusters, and delicate soft X-ray filaments in the Virgo cluster (fig. 23) all show that the central intracluster medium in these cool core clusters is not wildly turbulent. There are gas motions but they appear to be quite ordered on scales of several kpc. In the case of the Perseus cluster, horseshoe-shaped filaments trail behind a buoyant outer bubble in a manner similar to that observed in rising bubbles in the laboratory (fig. 30). This suggests that the viscosity may be high enough to dissipate the sound waves.

So far I have implied that the radiative cooling is completely shut off. Exactly how this happens is puzzling since the coolest gas at, say, $5 \times 10^6$ K has the shortest radiative cooling time of $<10^8$ yr and is spatially situated in blobs near, but not at the centre. Heating needs to target this gas without overheating its surroundings. No O vii emission lines have so far been seen, which means that there is no evidence for X-ray emitting gas below about $3 \times 10^6$ K. There is in many of these clusters considerable masses of both warm and cold molecular gas, seen either as $H_2$ or CO (Edge 2001, 2002, Salome *et al.* 2006). The total molecular mass of the filamentary system around NGC1275 in the Perseus cluster is $10^{11} M_\odot$. This assumes a conversion factor from CO measurements due to the successful internal heating model of Ferland *et al.* (2008).

Many of the cool core clusters with short central radiative cooling times have large masses of cold gas and dust. In some there is considerable star formation too; the elliptical galaxy with the highest star formation rate at low redshifts – the central galaxy in A1835 – has a star formation rate of about $100 M_\odot$ per year (O'Dea *et al.* 2008).

It is possible that non-radiative cooling is responsible. If the X-ray coolest clumps mix with the cold dusty molecular gas then, like milk in hot coffee, the temperature drops rapidly. There is certainly more than enough IR emission seen from these objects to account for this. When the cold gas is taken into account the heating–cooling balance is not necessarily so finely tuned. The balancing act is not eliminated but it is not quite so extreme.

### In conclusion

We have seen that accreting black holes can do considerable work on their surroundings. We now have observations of their effect from the innermost parts of the flow to the outer reaches of the most massive galaxies. In relative terms this is comparable to the size of a grape to the size of the Earth. Many fascinating problems are emerging which will keep us busy for a long time. ●

*A C Fabian, Institute of Astronomy, Madingley Road, Cambridge CB3 0HA, UK. With thanks to my many collaborators.*

### References

**Allen S** *et al.* 2006 *MNRAS* **372** 21.
**Brenneman L W and Reynolds C S** 2008 *ApJ* **652** 1028.
**Cackett** *et al.* 2008 *ApJ* **674** 415.
**Dabrowski Y** *et al.* 1997 *MNRAS* **288** L11.
**Dunn R J H and Fabian A C** 2006 *MNRAS* **373** 959.
**Edge A C** 2001 *MNRAS* **328** 762.
**Edge A C** *et al.* 2002 *MNRAS* **337** 49.
**Elvis M** *et al.* 2002 *ApJ* **565** L75.
**Fabian A C** 1999 *MNRAS* **308** L39.
**Fabian A C and Miniutti** 2009 in *The Kerr Spacetime* ed. Wiltshire *et al.* (Cambridge University Press).
**Fabian A C** *et al.* 1989 *MNRAS* **238** 729.
**Fabian A C** *et al.* 2002 *MNRAS* **329** L18.
**Fabian A C** *et al.* 2003 *MNRAS* **344** L43.
**Fabian A C** *et al.* 2006 *MNRAS* **366** 417.
**Fabian A C** *et al.* 2008 *Nature* **454** 968.
**Fabian A C** *et al.* 2009a *MNRAS* **394** L89.
**Fabian A C** *et al.* 2009b *Nature* May 28.
**Ferland G J** *et al.* 2008 *MNRAS* **392** 1475.
**Forman W** *et al.* 2007 *ApJ* **665** 1057.
**George I M and Fabian A C** 1991 *MNRAS* **249** 352.
**Guainazzi M** 1999 *A&A* **341** L27.
**Guainazzi M** *et al.* 2006 *AN* **327** 1032.
**King A R and Pringle J E** 2007 *MNRAS* **377** L25.
**Laor A** 1991 *ApJ* **376** L90.
**Lightman A P and White T R** 1988 *ApJ* **335** L57.
**Marconi A** *et al.* 2004 *MNRAS* **351** 169.
**McNamara B R and Nulsen P E J** 2007 *ARAA* **45** 117.
**Miller J M** 2007 *ARAA* **45** 441.
**Miller J M** *et al.* 2008 *ApJ* **679** L113.
**Miller J M** *et al.* 2009 *ApJ* **697** 900.
**Miller L** *et al.* 2008 *A&A* **483** 437.
**Miniutti G and Fabian A C** 2004 *MNRAS* **349** 1435.
**Miniutti G** *et al.* 2007 *PASJ* **59** 315.
**Nandra K** *et al.* 2008 *MNRAS* **382** 194.
**O'Dea C P** *et al.* 2008 *ApJ* **681** 1035.
**Peterson J R and Fabian A C** 2006 *PhR* **427** 1.
**Raimundo S and Fabian A C** 2009 *MNRAS* in press (arXiv:0903.3432).
**Reis R C** *et al.* 2008 *MNRAS* **387** 1489.
**Reynolds C S** *et al.* 2009 *MNRAS* in press (arXiv:0904.3099).
**Ross R R and Fabian A C** 1993 *MNRAS* **261** 74.
**Ross R R and Fabian A C** 2005 *MNRAS* **358** 211.
**Salome P** *et al.* 2006 *A&A* **454** 437.
**Sanders J S** *et al.* 2007 *MNRAS* **385** 1186.
**Shafee R** *et al.* 2006 *ApJ* **636** L113.
**Silk J and Rees M J** 1998 *A&A* **331** 1.
**Soltan A** 1982 *MNRAS* **200** 115.
**Tanaka Y** *et al.* 1995 *Nature* **375** 659.
**Thorne K S** 1974 *ApJ* **191** 507.
**Volonteri M** *et al.* 2005 *ApJ* **620** 96.
**Yu Q and Tremaine S** 2002. *ApJ* **335** 965.